\documentclass[10pt]{article}
\usepackage{fullpage}
\usepackage{amsmath}
\usepackage{amsfonts}
\usepackage{epsfig}

\def\##1{{\bf #1}}
\def\=#1{\underline{\underline{#1}}}

\def\+#1{\underline{\bf #1}}
\def\*#1{\underline{\underline{\bf #1}}}

\def\c#1{\cite{#1}}

\def\lec{\left\{}
\def\ric{\right\}}

\def\.{\mbox{ \tiny{$^\bullet$} }}

\begin{document}

\begin{center}

{\bf {\LARGE A Plethora of Negative-Refraction Phenomenons in
      Relativistic
      and Non-Relativistic Scenarios}}

\vspace{5mm}

{\bf {\large Tom G. Mackay${}^{a,b,}$\footnote{Email: T.Mackay@ed.ac.uk} and Akhlesh Lakhtakia${}^{b,}$\footnote{Email: akhlesh@psu.edu} }}\\

\vspace{5mm}

${}^{a}$School of Mathematics and
   Maxwell Institute for Mathematical Sciences,\\
University of Edinburgh, Edinburgh EH9 3JZ, UK \\
${}^{b}$NanoMM---Nanoengineered Metamaterials Group, Department of Engineering Science and Mechanics,\\
Pennsylvania State University, University Park, PA 16802-6812, USA

\end{center}
\vspace{5mm}

\begin{abstract}
In accordance with Snel's law of refraction, whether a plane wave is
refracted in the negative sense or positive sense at a planar
boundary between two homogenous mediums is determined solely by the
orientation of the real parts of the wavevectors involved. Thus,
negative refraction should be distinguished from the associated but
independent phenomenons of negative phase velocity, counterposition
and negative deflection of energy flux. None of these phenomenons is
Lorentz covariant.
\end{abstract}

\vspace{20mm}

\section{Introduction}

Despite optical refraction being one of the oldest topics of
scientific investigation,  interest in the phenomenon of negative
refraction has been widespread only for the past ten years. This interest
stemmed from experimental reports of negative refraction, at microwave
frequencies, in certain carefully engineered metamaterials
\c{Shelby}. More recently, negative refraction in metamaterials has
been reported at higher frequencies, with the visible frequency
range now almost attained \c{Shalaev}. Research on negative
refraction has been largely fuelled by the prospect of manufacturing
planar lenses  from negatively refracting materials \c{Pendry}.
While these lenses would not be `perfect' \c{Ziolkowski,Lakh2002}, they may
have a very high resolving power.

The present--day  scientific literature on optics abounds with
descriptions of purported negative refraction in a variety of
complex scenarios. However,   some inconsistencies are apparent
between reports over what exactly constitutes negative refraction.
Sometimes what is actually being described is not negative
refraction \emph{per se}, but an associated phenomenon such as
negative phase velocity, counterposition or negative deflection of
energy flux \c{Jen2009,ML2009_PRB}. While these phenomenons
can---and often do---arise in conjunction with negative
refraction, it is important to bear in mind that these are
independent phenomenons which should be distinguished from negative
refraction, especially so in complex materials and in relativistic
scenarios.

We highlight here the distinctions between
negative refraction and the associated phenomenons of negative phase
velocity, counterposition and negative deflection of energy flux.
These distinctions are apparent in both relativistic and
non-relativistic scenarios.

\section{Negative refraction}

\begin{figure}[h]
    \centerline{\psfig{figure=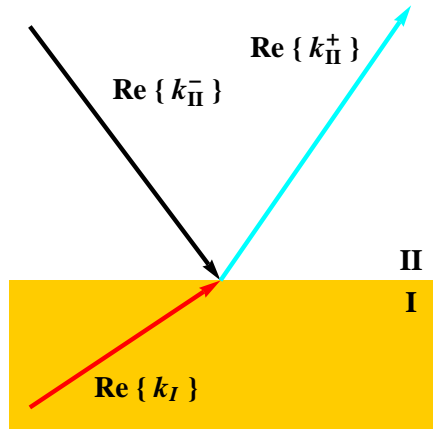,width=58.7mm} }
    \caption{A schematic illustration of  refraction of a plane wave at the planar interface between mediums I and II.
     Suppose a uniform plane wave propagates in medium I with wavevector $\#k_I$, towards the interface.
      If it is negatively refracted at the interface then its wavevector in medium II,
      namely $\#k^-_{II}$, is such that $\mbox{Re} \lec \#k^-_{II} \ric
       \. \hat{\#n} < 0$, where $\hat{\#n}$ is the unit vector normal to the interface directed into medium II.
      If it is positively refracted at the interface then its wavevector in medium II,
      namely $\#k^+_{II}$, is such that $\mbox{Re} \lec \#k^+_{II} \ric \. \hat{\#n} > 0$.}
    \label{schematic_nr}
\end{figure}

Suppose a  plane wave propagates in medium I with wavevector
$\#k_I$, towards a planar interface separating medium I and medium
II, such that $\mbox{Re} \lec \#k_I  \ric \. \hat{\#n} > 0$, where
$\hat{\#n}$ is the unit vector normal to the interface directed into
medium II. In keeping with Snel's law of refraction \c{Jackson}, the
plane wave is said to be negatively refracted at the planar
interface between mediums I and II, if the dot product $\mbox{Re}
\lec \#k_{II} \ric \. \hat{\#n} < 0$ where $\#k_{II}$ is the wave
vector in medium II. Conversely, if $\mbox{Re} \lec \#k_{II} \ric \.
\hat{\#n} > 0$ then the plane wave is said to be positively
refracted. A schematic illustration is provided in
Fig.~\ref{schematic_nr}.

In the case of uniform plane waves propagating in passive isotropic
dielectric--magnetic materials, negative refraction is synonymous
with negative phase velocity. However, the correspondence between
negative refraction and negative phase velocity breaks down if the
materials are active \c{LMG2009_MOTL}, or if nonuniform plane waves
are considered \c{ML2009_PRB}, or if anisotropic or bianisotropic
materials are involved \c{ML2009_PRB}.

Furthermore, whether a plane wave is refracted positively or negatively, depends
upon the inertial frame of reference of the observer, as was
recently demonstrated using a pseudochiral omega material moving at
uniform velocity \c{ML2009_PLA}.

\section{Negative phase velocity}

\begin{figure}[h]
\centerline{
\psfig{figure=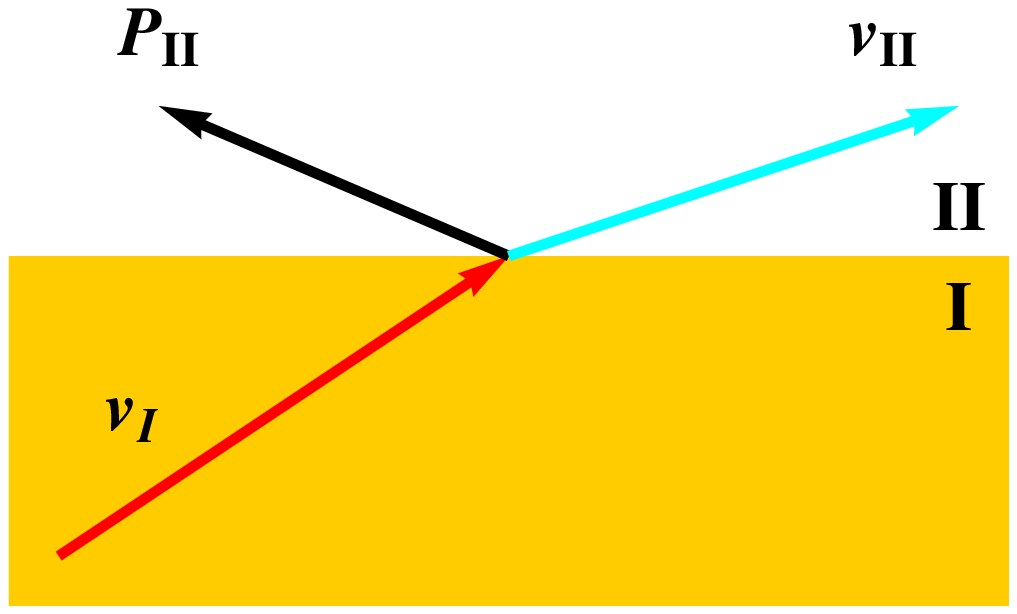,width=40.7mm} \\
\psfig{figure=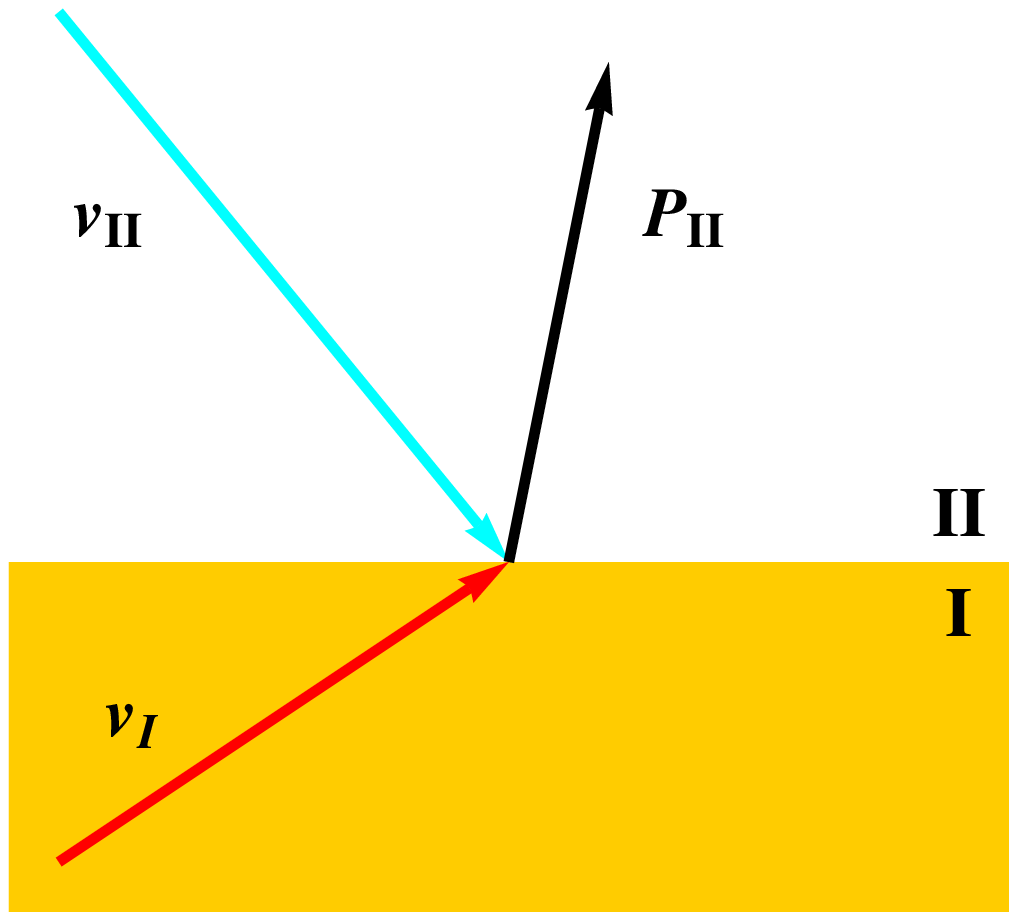,width=40.7mm}}
    \caption{Two schematic examples of negative phase velocity, arising  at the planar interface between mediums I and II.
     Suppose a plane wave propagates in medium I with phase velocity $\#v_I$, towards the interface.
     The corresponding plane wave in medium II has negative phase velocity if $\#v_{II} \. \#P_{II} < 0 $,
where $\#v_{II}$ and $\#P_{II}$ are the phase velocity  and
time--averaged Poynting vector, respectively, of the plane wave in
medium II. In the example on the left side, the plane wave in medium
II is positively refracted and the wavevector and time--averaged
Poynting vector are counterposed. In the example on the right side,
the plane wave in medium II is negatively refracted and  the
wavevector and time--averaged Poynting vector are not counterposed.}
    \label{schematic_npv}
\end{figure}

The sign associated with a plane wave's velocity refers to the direction of the
phase velocity vector relative to the corresponding time--averaged
Poynting vector.
 A plane
wave with phase velocity $\#v$ and time--averaged Poynting vector
$\#P$ has negative phase velocity if $\#v \. \#P < 0$ and positive
phase velocity if $\#v \. \#P > 0$. We note that orthogonal phase
velocity, i.e., $\#v \. \#P = 0$, is also a possibility
\c{ML2009_PRB}. Although negative phase velocity and negative
refraction often accompany each other, they need not. The schematic
Fig.~\ref{schematic_npv} shows how negative phase velocity can arise
with or without negative refraction.

While the simplest material that can support negative refraction is
an isotropic dielectric--magnetic material, it is notable that even
an isotropic dielectric material can support negative phase velocity
(in conjunction with positive refraction) \c{ML2009_PRB}. However,
there is greater scope for negative phase velocity in more complex
materials such as isotropic  chiral  \c{M05_MOTL,ML2010_SPIE_Rev} and
bianisotropic \c{ML04_PRE,ML2005_NJP} materials.

As is the case for refraction, the sign of a plane wave's phase
velocity is not Lorentz covariant \c{ML2009_JPA}.\footnote{A
covariant analogue of the NPV condition $\#k \. \#P < 0$ has been
derived, where $\#k$ is a real--valued wavevector \c{McCall_PRL}.
However, the physical basis for this is unclear because there is no
\emph{a priori} reason for the quantity $\#k \. \#P$ to be
covariant.} A noncovariant formalism---wherein vacuum in curved
spacetime is formally represented by a fictitious nonhomogeneous
bianisotropic medium in flat spacetime \c{Schleich}---has been
widely used to investigate the prospects of negative phase velocity
in vacuum under the influence of strong gravitational fields
\c{LM2004_JPA,LMS2005_PLA,MLS2005_NJP}. The metrics associated with
de Sitter \c{MSL2005_EPJC}, Kerr \c{Komissarov, Sharif},
Kerr--Newman \c{Optik}, Reissner--Nordstr\"om \c{Ali_Hasan},
Schwarzschild \c{Sharif_GRG}, and Schwarzschild--de Sitter
\c{MLS2005_EPL} spacetimes have been found to support negative phase
velocity. Let us note that negative phase velocity in such general
relativistic scenarios is quite distinct from superradiance,
although both phenomenons involve negative energy densities
\c{Superradiance}.

\section{Counterposition}

\begin{figure}[h]
\centerline{\psfig{figure=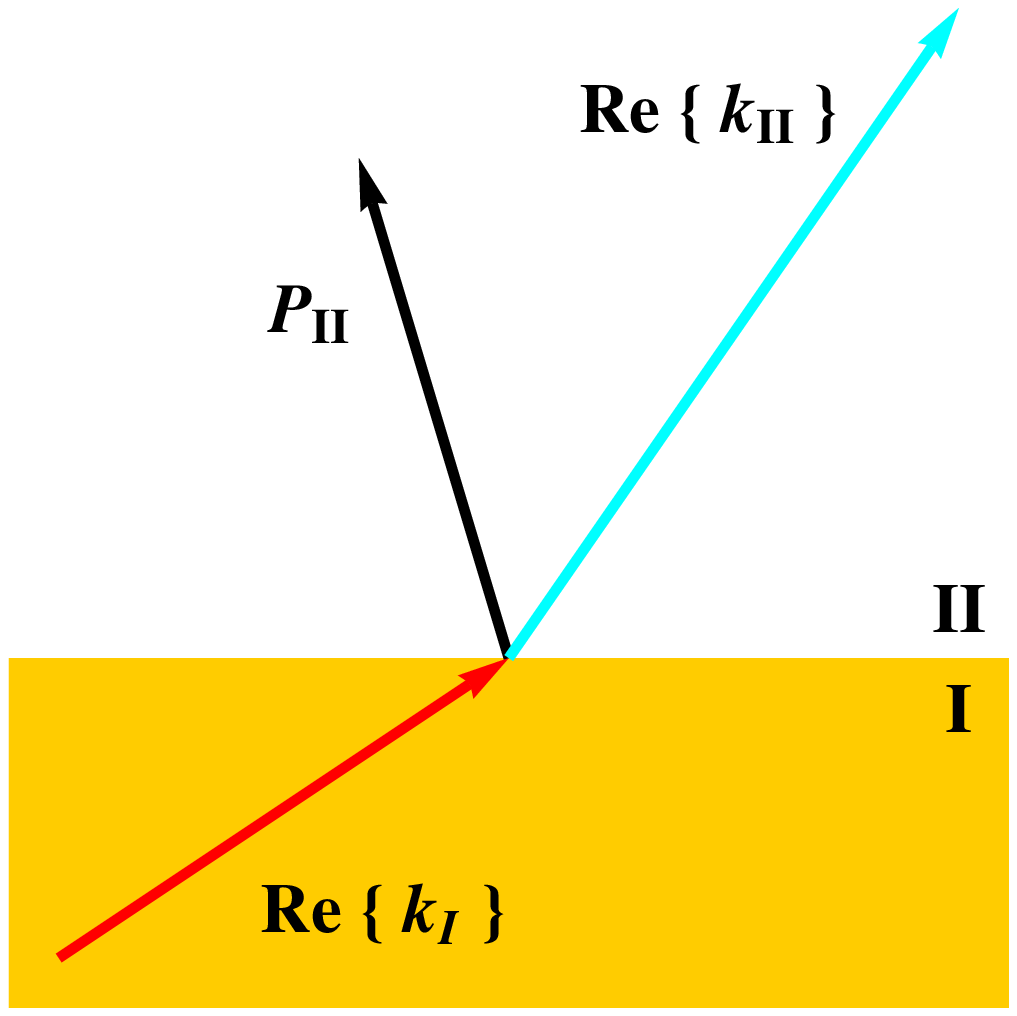,width=40.7mm}\\
\psfig{figure=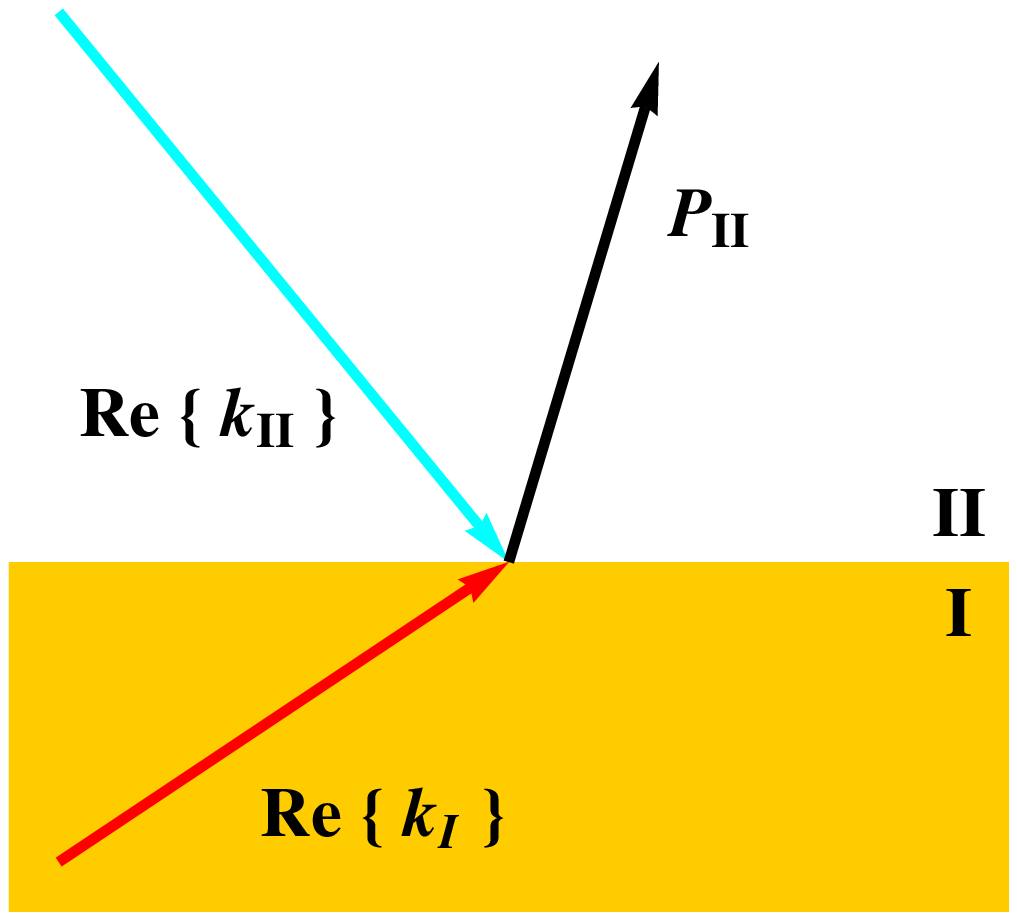,width=40.7mm}}
    \caption{Two schematic examples of counterposition, arising at the planar interface between mediums I and II.
     Suppose a  plane wave propagates in medium I with wavevector $\#k_I$, towards the interface.
     The corresponding real part
     of the wavevector $\#k_{II}$ and time--averaged Poynting vector $\#P_{II}$ in medium II are counterposed if they lie on opposite sides of the
     the unit vector normal to the interface directed into medium II. In the example on the left side, the plane wave in medium II is
     positively refracted and has positive
      phase velocity. In the example on the right side, the plane wave in medium II is negatively refracted and has negative phase velocity.}
    \label{schematic_cp}
\end{figure}

Consider a  plane wave, propagating in medium I towards a planar
interface between mediums I and II. The real part of the wavevector
and the time--averaged Poynting vector of the resulting  plane wave
which is launched in medium II are said to be counterposed if these
two vectors lie on opposite sides of the unit vector normal to the
interface between mediums I and II. A schematic illustration of two
different manifestations of counterposition---one arising in
conjunction with negative refraction and the other not---is
provided in Fig~\ref{schematic_cp}.

Counterposition cannot occur in isotropic dielectric--magnetic
materials (at rest), but it can in anisotropic materials
\c{LM04_Optik} where it is sometimes referred to as amphoteric
refraction \c{Zhang}. Indeed, counterposition has also been referred
to as negative refraction \c{Kong}. Similarly to negative refraction
and negative phase velocity, counterposition can be induced by
motion at constant velocity in materials which do not support
counterposition at rest \c{ML2009_JPA,Kong,ML2007_MOTL}.

\section{Negative deflection of energy flux}

\begin{figure}[h]
    \centerline{\psfig{figure=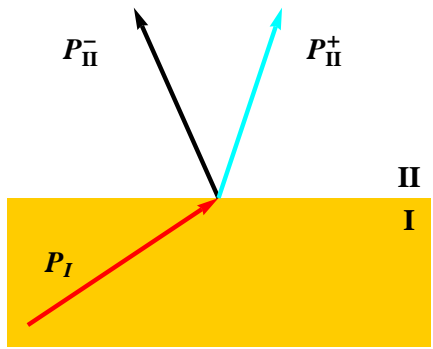,width=58.7mm} }
    \caption{A schematic illustration of  deflection of a energy flux of a plane wave, at the planar interface between mediums I and II.
     Suppose a plane wave propagates in medium I with time--averaged Poynting vector $\#P_I$, towards the interface, such that $\#P_I \. \hat{\#t} > 0$
      where $\hat{\#t}$ is a unit vector lying along the interface between mediums I and II, in the plane of incidence.
      If the plane wave's energy flux is  negatively deflected at the interface then the corresponding
      time--averaged Poynting vector in medium II,
      namely $\#P^-_{II}$, is such that $\#P^-_{II} \. \hat{\#t} < 0$.
      If it is positively deflected at the interface then the corresponding
      time--averaged Poynting vector in medium II,
      namely $\#P^+_{II}$,
is such that $\#P^+_{II} \. \hat{\#t} > 0$.}
    \label{schematic_pvector}
\end{figure}

Although there is no equivalent of Snel's law for energy flux, the
notion of negative deflection of energy flux has been adopted by
some in the negative refraction community
\c{Zhang,Belov,Shen,Hoffman,Yao}. The energy flux associated with a
plane wave---as represented by the time--averaged Poynting
vector---is negatively deflected at a planar boundary between
mediums I and II if the time--averaged Poynting vectors in mediums I
and II lie on  opposite sides of the unit vector normal to the
interface. A schematic illustration of negative and positive energy
flux deflection is provided in Fig.~\ref{schematic_pvector}. For
uniform plane waves in isotropic dielectric--magnetic materials, the
direction of  energy flux is parallel or anti--parallel to the
wavevector, but for more complex materials and/or nonuniform plane
waves  this is not the case
 \c{ML2009_PRB}. Thus, the energy  flux can deflected negatively even
 if though the plane wave is refracted positively, and vice versa.
Furthermore, whether or not the energy flux is deflected negatively
depends upon the inertial reference frame of the observer
\c{ML2009_JPA}.

From a practical point of view, the direction of energy flow
associated with a beam is likely to be of greater significance than
the energy flow associated with a plane wave. Of course, many
standard optical devices can be straightforwardly implemented to
deflect beams in any direction, without the need for complex
materials or metamaterials, and it is also possible to suppress
beam deflection \c{Shifrin,Bohren}. A recent theoretical study
involving beam propagation through uniformly moving slabs has
demonstrated that whether a beam is negatively or positively
deflected depends upon the  motion of the slab relative to an
observer \c{ML2009_JPA}. This result could be harnessed to achieve a
degree of concealment: if a slab moves at a certain specific
velocity relative to an observer, a beam can propagate through the
slab with no deflection at all \c{ML2007_concealment}.

\section{Closing remarks}

Negative refraction is already the focus of  considerable research
efforts and it may well play  significant roles in future optical
technologies. It is therefore important to unambiguously establish
exactly what is negative refraction, and to distinguish it from the
associated but independent phenomenons of negative phase velocity,
counterposition and negative deflection of energy flux. These
distinctions  are of particular importance when more complex
materials and metamaterials are considered, and in relativistic
scenarios.

\section*{Acknowledgment}
TGM is supported by a  Royal Academy of Engineering/Leverhulme Trust
Senior Research Fellowship. AL thanks the Charles Godfrey  Binder
Endowment at Penn State for partial financial support of his
research activities.

\end{document}